\documentstyle[12pt]{article}
\newcommand{\beq}{\begin{equation}}
\newcommand{\eeq}{\end{equation}}
\newcommand{\bea}{\begin{eqnarray}}
\newcommand{\eea}{\end{eqnarray}}
\newcommand{\tr}{{\,\hbox{\rm Tr}\,}}
\newcommand{\one}{{\hbox{\bf 1}}}
\newcommand{\zero}{{\hbox{\bf 0}}}
\newcommand{\la}{\left\langle}
\newcommand{\ra}{\right\rangle}

\newcommand{\bfe}{{\hbox{\bf e}}}
\newcommand{\bfp}{{\hbox{\bf p}}}
\newcommand{\bfr}{{\hbox{\bf r}}}
\newcommand{\bfR}{{\hbox{\bf R}}}
\newcommand{\bfv}{{\hbox{\bf v}}}
\newcommand{\bfphi}{{\vec\phi}}

\newcommand{\calf}{{\cal F}}
\newcommand{\calj}{{\cal J}}
\newcommand{\caln}{{\cal N}}

\newcommand{\barphi}{{\bar \phi}}

\newcommand{\de}{{\,\cal D}}

\newcommand{\rf}[1]{{(\ref{#1})}}
\newcommand{\expo}[1]{{\exp\left(#1\right)}}

\begin{document}
\begin{titlepage}
\begin{flushright}
July 1995\\
\end{flushright}
\vspace{0.5cm} 
\begin{center} 
{\Large{\bf 
The Lyapunov Spectrum of a Continuous \\ 
Product of Random Matrices
}} \\
\vspace{2cm}
{\bf Andrea Gamba}
\footnote{E-mail: gamba@pol88a.polito.it} \\
\vspace{0.4cm}
{\em Dipartimento di Matematica, Politecnico di Torino, 10129 Turin, 
Italy \\
and INFN, sezione di Milano, 20133 Milan, Italy} \\
\vspace{0.4cm}
{\bf Igor V. Kolokolov}
\footnote{E-mail: kolokolov@vxinpb.inp.nsk.su} \\
\vspace{0.4cm}
{\em Budker Institute of 
Nuclear Physics, 630090 Novosibirsk, Russia.} \\

\end{center}
\vspace{2cm}

\begin{abstract}
\noindent
We expose a functional integration method
for the averaging of continuous products $\hat{P}_t$
of $N\times N$ random matrices.
As an application, we compute exactly the statistics
of the Lyapunov spectrum of $\hat{P}_t$.
This problem is relevant to the study of the statistical 
properties of various
disordered physical systems, and specifically
to the computation of the multipoint correlators of a 
passive scalar
advected by a random
velocity field. 
Apart from these applications,
our method provides a general setting for computing statistical
properties of linear evolutionary systems subjected to a white
noise force field.
\\ \phantom{.}
\end{abstract}

\end{titlepage}
\newpage
\baselineskip=18pt

\section{Introduction}

In this work we give a detailed exposition of a functional
integral method  for the averaging of time-ordered exponentials
of $N\times N$ random matrices which has found several applications
in the study of the statistical properties  of disordered systems.

The method was introduced by one of the authors~\cite{k1} in the
$N=2$ case in order to compute the partition function
of the Heisenberg ferromagnet, and was thereafter applied to
the study of one-dimensional Anderson localization~\cite{alpha}
and to some problems of mesoscopic physics~\cite{beta}.
Later, the same technique~\cite{ps1,ps} allowed to obtain analytical
results in the problem of a passive scalar advected by a 2-dimensional
random velocity field.
The approach of refs.~\cite{ps1,ps} was then extended to the more general
$N$-dimensional case in ref.~\cite{chgk}.

In the present work we expose the method in its full generality
and show that it allows to compute exactly the statistics
of the whole Lyapunov spectrum of the matrix $\hat{P}_t$
describing the time evolution of a linear system subjected
to a white noise force field.
Such a spectrum is relevant~\cite{ps2} in the computation
of the multipoint correlators of the passive scalar
(as well as in the computation of
the correlators of passive vectors and tensors);
in this case the matrix $\hat{P}_t$ describes the time
evolution of particles in a turbulent fluid,
linearized around a given trajectory. 

However, our setting presents a high degree of generality and
therefore a wider range of applications. 
In particular, our formalism can be naturally applied to the study of 
$N$-level quantum-mechanical  systems affected by a random noise,
and 
it is complementary
to the supersymmetric approach
~\cite{su1,su2} to the problem of $N/2$ channels localization 
in a disordered conductor.

\section{Averages of time-ordered exponentials}

Let us start with the problem of computing gaussian averages
of the form
\beq\label{uno}
\la \calf[\hat{P}_t] \ra = 
{1\over \caln}\int \de \hat{X} \exp 
\left(
-S[\hat{X}]
\right) 
\, \calf[\hat{P}_t]
\eeq
where
$\hat{X}(s)$, for
$0\le s\le T$,
is a traceless $N\times N$ 
hermitian matrix,
\beq\label{unobis}
\de \hat{X}\equiv\prod_{0\le s\le T}\prod_{i<j}
dX_{ij}(s)dX_{ji}(s)\prod_i dX_{ii}(s)
\eeq
is the Feynman-Kac measure, $\caln$ 
is chosen in such a way that $\la 1\ra=1$,
\beq\label{unoter}
S[\hat{X}]={1\over 4D} \int_0^T\tr\hat{X}^2(s)\,ds
\eeq
and $\hat{P}_t$ is the time-ordered exponential
\beq\label{due}
\hat{P}_t={\cal T}\exp
\left(
\int_0^t \hat{X}(s)\,ds
\right)
\eeq
such that 
$\bfr(t)=\hat{P}_t\bfr_0$
is the general solution of the linear problem
$\dot\bfr=\hat{X}\bfr$, $\bfr(0)=\bfr_0$
and 
\beq\label{tre}
\dot{\hat{P}}_t\hat{P}^{-1}_t=\hat{X}
\eeq 

We shall now introduce a set of ``collective'' integration variables
in order to re-express~\rf{due} in a more tractable form. 
At the same time, we shall chose the new variables in such a way 
that~\rf{unoter} still be quadratic and that the Jacobian
determinant of the functional transformation  be particularly simple.

As a first step, let us Gauss-decompose the matrix $\hat{P}_t$:
\beq\label{quattro}
\hat{P}_t=
(\one+\hat{\phi}(t))\cdot\exp(\hat{\tau}(t))\cdot(\one+\hat{\theta}(t))
\eeq
where
\bea
\phi_{ij}(t)	&\equiv&	 0,\qquad i\le j 	\nonumber	\\
\theta_{ij}(t)	&\equiv&	 0,\qquad i\ge j 	\label{cinque} 	\\
\tau_{ij}(t)	&\equiv&	 \tau_i(t)\,\delta_{ij}	\nonumber	\\
\tau_N(t)	&\equiv&   -\sum_{j=1}^{N-1}\tau_j(t)	\label{somma}
\eea
Moreover, in order to ensure the equality 
$\hat{P}_0=\one$
we shall impose
\beq\label{cinquebis}
\hat{\phi}(0)=\zero,\qquad \hat{\theta}(0)=\zero
\eeq
We would like now to re-express the ``local'' degrees of freedom
$X_{ij}(t)$
in terms of the global ones 
${\phi}_{ij}(t),{\tau}_{ij}(t),{\theta}_{ij}(t)$.
This can be accomplished by making use of the basis $\hat{e}_{ij}$
of the matrix algebra, which is defined by
$(\hat{e}_{ij})_{kl}=\delta_{ik}\delta_{jl}$
and satisfies the commutation rules
\beq\label{sette}
[\hat{e}_{ij},\hat{e}_{kl}]=\delta_{jk}\hat{e}_{il}-\delta_{il}\hat{e}_{kj}
\eeq
In particular, one has
\bea\nonumber
&&\hat{e}_{ii}\hat{e}_{kl}=\hat{e}_{kl}(\hat{e}_{ii}+\delta_{ik}-\delta_{il})
\\
\label{comm}
&&e^{\hat{\tau}}\,\hat{e}_{ij}\,e^{-\hat{\tau}}
=e^{\tau_i-\tau_j}\,\hat{e}_{ij}
\eea
From these relations  the desired expression for $X_{ij}$
readily follows:
\bea\nonumber
X_{ij}	&=&
\dot{\phi}_{ij}+\sum_k\dot{\phi}_{ik} \tilde{\phi}_{kj}
\\
\label{sei}
&&+\dot{\tau}_{i}\delta_{ij}+\phi_{ij}\dot{\tau}_j
+\dot{\tau}_i\tilde{\phi}_{ij}+\sum_k\phi_{ik}\dot{\tau}_k\tilde{\phi}_{kj}
\\
&&+A_{ij}+\sum_k(\phi_{ik}A_{kj}+A_{ik}\tilde{\phi}_{kj})
+\sum_{k,l}\phi_{ik}A_{kl}\tilde{\phi}_{lj}
\nonumber
\eea
where 
$A_{ij}\equiv e^{\tau_i-\tau_j}\sum_k\dot{\theta}_{ik}(\delta_{kj}+
\tilde{\theta}_{kj})$,
\beq\label{nove}
\tilde\phi_{ij}
\equiv
-\phi_{ij}+
\sum_k\phi_{ik}\phi_{kj}
-\sum_{k,l}\phi_{ik}\phi_{kl}\phi_{lj}+\cdots
\eeq
and a similar definition holds for $\tilde\theta_{ij}$ 
(for any fixed $N$~\rf{nove} is a finite sum,
since $\hat{\phi}$ is a nilpotent matrix; the same is true for $\hat{\theta}$).

Substituting~\rf{sei} in~\rf{unoter} one obtains
\beq\label{otto}
{1\over 2}\tr\hat{X}^2 = {1\over 2} \sum_{j=1}^N\dot\tau_j^2 
+\sum_{i,j}
\left(
	\dot\phi_{ij}+\sum_{k=1}^N\tilde\phi_{ik}\dot\phi_{kj}
\right)
e^{\tau_j-\tau_i}
\left(
	\sum_{k=1}^N\dot\theta_{jl}\tilde\theta_{li}+\dot\theta_{ji}
\right)
\eeq
The form of~\rf{otto} suggests the introduction of the new variables
\bea\label{dieci}
\barphi_{ij}
&=&
\sum_{k,l}\dot\theta_{ik}\left(\delta_{kl}+\tilde\theta_{kl}\right)
\left(\delta_{lj}+\tilde\phi_{lj}e^{\tau_j-\tau_l}\right)
e^{\tau_i-\tau_j},
\qquad	i<j
\\
\barphi_{ij}&\equiv& 0,	\qquad	i\ge j
\nonumber
\eea
so that
\beq\label{undici}
{1\over 2}\tr\hat{X}^2
=
{1\over 2}\sum_{j=1}^N\dot\tau_j^2
+\sum_{i,j}\dot\phi_{ij}\barphi_{ji}
\eeq
Relation~\rf{dieci} can be inverted giving
\beq\label{dodici}
\dot\theta_{ij}=\sum_{k,l}
\barphi_{ik}
e^{\tau_k-\tau_i}
\left(\delta_{kl}+\phi_{kl}e^{\tau_l-\tau_k}\right)
\Big(\delta_{lj}+\theta_{lj}\Big)\chi_{li}
\eeq
where
\beq\label{tredici}
\chi_{ij}\equiv 1-\bar\chi_{ij}\equiv
\left\{
\matrix{1,&\quad i>j \cr 0,&\quad i\le j}
\right.
\eeq
Through~\rf{dodici} one can re-express the $\theta_{ij}$ as functions
of the new variables $\phi_{ij},\tau_i$ and $\bar\phi_{ij}$
in a recursive way, thanks to the ``triangular'' form of the equation.
For instance, 
for $N=3$
one gets 
\bea
\theta_{23}(t)	&=&	\int_0^t\bar\phi_{23}(s)
e^{-\tau_1(s)-2\tau_2(s)}\,ds,
\nonumber\\
\theta_{12}(t)	&=&	\int_0^tA(s)\,ds,\qquad\hbox{where}\quad
A=(\bar\phi_{12}+\bar\phi_{13}\phi_{32})e^{\tau_2-\tau_1}\, ,
\label{quattordici}\\
\theta_{13}(t)	&=&	\int_0^t
[A(s)\theta_{23}(s)+\bar\phi_{13}(s)e^{-2\tau_1(s)-\tau_2(s)}]\,ds
\nonumber
\eea
As a matter of fact, for any fixed $i$ the $N-i$ functions $\theta_{ij}$
can be expressed through the $N-i-1$ functions $\theta_{i+1,j}$ and the
remaining variables by means of a single quadrature.
This is an important point, since for practical calculations
$\hat{P}_t$ has to be re-expressed in terms of the new variables
$\hat{\phi},\dot{\hat{\tau}},\hat{\bar\phi}$.

We must now substitute $\hat{X}(s)$ as an integration variable in the
functional integral~\rf{uno} with the new variables 
$\hat{\phi}(s),\dot{\hat{\tau}}(s),\hat{\bar\phi}(s)$.
Again using the commutation rules~\rf{sette}, and renaming 
$\dot{\tau}_i\equiv{\rho}_i$
for convenience, we finally get
\bea
X_{ij}	&=&	\bar\phi_{ij}+\sum_{k}\phi_{ik}\bar\phi_{kj},
\qquad i<j
\nonumber\\
X_{ii}	&=&	\rho_i+
\sum_{k}(\phi_{ik}\bar\phi_{ki}-\bar\phi_{ik}\phi_{ki}),
\qquad i=1,\ldots,N
\\
\label{quindici}
X_{ij}	
&=&	\phi_{ij}\rho_j+\rho_i\tilde\phi_{ij}+\dot\phi_{ij}
\\
&+&	\sum_{k}
(\phi_{ik}\rho_k\tilde\phi_{kj}+\dot\phi_{ik}\tilde\phi_{kj}
+\phi_{ik}\bar\phi_{kj}-\bar\phi_{ik}\phi_{kj})	
\nonumber\\
&-&	\sum_{k,l}(\bar\chi_{jk}\phi_{ik}\bar\phi_{kl}\phi_{lj}
+\bar\chi_{li}\bar\phi_{ik}\phi_{kl}\tilde\phi_{lj})
\nonumber\\
&-&	\sum_{k,l,m}\bar\chi_{mk}\phi_{ik}\bar\phi_{kl}\phi_{lm}\tilde\phi_{mj},
\qquad i>j 
\nonumber
\eea
In ref.~\cite{chgk} the $N=3$ case was explicitly considered.
We observe that the matrix elements of $\hat{X}(t)=\dot{\hat{P}}_t\hat{P}_t^{-1}$
transform as
\beq\label{sedici}
X_{ij}(t)\rightarrow e^{\sigma_{ij}}X_{ij}(t)
\eeq
under the global gauge transformation
\beq\label{diciassette}
\phi_{ij}(t)\rightarrow e^{\sigma_{ij}}\phi_{ij}(t),	\qquad
\bar\phi_{ij}(t)\rightarrow e^{\sigma_{ij}}\bar\phi_{ij}(t),	\qquad
\rho_i(t)\rightarrow\rho_i(t)
\eeq
where $\sigma_{ij}$ satisfies 
$\sigma_{ik}+\sigma_{kj}=\sigma_{ij}$,
$\sigma_{ij}=-\sigma_{ji}$.

As the last step, we must compute the Jacobian determinant of the
functional transformation~\rf{quindici}.
Notice first that through the shift given by
\bea\label{diciotto}
\bar\phi_{ij}	&\rightarrow&	\barphi_{ij}
-\sum_{k=1}^N\phi_{ik}\bar\phi_{kj}
\\
\rho_i	&\rightarrow&	\rho_i
-\sum_{k=1}^N(\phi_{ik}\bar\phi_{ki}-\bar\phi_{ik}\phi_{ki})
\nonumber
\eea
(which has Jacobian $\calj'=1$)
one can reduce to the computation of
\beq\label{diciannove}
\hbox{Det}\left(
{\delta\hat{X}_-
\over
\delta\hat{\phi}\,\delta\hat{\rho}\,\delta\hat{\bar\phi}}
\right)
\eeq
where $\hat{X}_-$ is the strictly lower triangular part of $\hat{X}$.
The Jacobian~\rf{diciannove} can be computed by means of the standard
regularization procedure (see ref.~\cite{k1})
\bea
\nonumber
&&
\hat{\phi}_n=\hat{\phi}(t_n),	\qquad
\hat{\rho}_n=\hat{\rho}(t_n),	\qquad
\hat{\bar\phi}_n=\hat{\bar\phi}(t_n)
\\
&&
t_n=hn,\quad n=1,\ldots,M,\quad h=T/M\rightarrow 0,\quad M\rightarrow +\infty
\label{venti}
\\
&&
\dot{\hat{\phi}}(t)\rightarrow{\hat{\phi}_n-\hat{\phi}_{n-1}\over h},\qquad
\hat{\phi}(t)\rightarrow{\hat{\phi}_n+\hat{\phi}_{n-1}\over 2}
\nonumber
\eea
giving
\beq\label{ventibis}
\calj\propto\expo{\sum_{j=1}^{N-1}(N-j)\int_0^T\rho_j(s)\,ds}
\eeq
Applying now the variable transformation 
$\hat{X}\rightarrow(\hat{\phi},\hat{\rho},\hat{\bar\phi})$
one sees that the functional integral~\rf{uno} reduces to
\beq\label{ventuno}
\la\calf[\hat{P}_t]\ra
={1\over{\cal N}'}\int\de\hat{\phi}\de\hat{\rho}\de\hat{\bar\phi}
\,\,\expo{-S'[\hat{\phi},\hat{\rho},\hat{\bar\phi}]}
\calf[(\one+\hat{\phi})e^{\hat{\tau}}(\one+\hat{\theta})]
\eeq
where
$\hat{\theta}=\hat{\theta}[\hat{\phi},\hat{\rho},\hat{\bar\phi}]$
is obtained by solving~\rf{dodici}, 
$\hat{\tau}=\int_0^T\hat{\rho}(s)\,ds$,
the $\rho_i$ are constrained by~\rf{somma},
${\cal N}'$ is the normalization factor
and
\beq\label{ventidue}
S'=
{1\over 2D}
\int_0^T
\left(
	{1\over 2}\sum_{k=1}^N\rho_k^2
	+\sum_{i,j}\dot\phi_{ij}\bar\phi_{ji}
	-2D\sum_{k=1}^{N-1}(N-k)\rho_k
\right)\,ds
\eeq
In~\rf{uno} the functional integration is constrained to the surface
\beq\label{ventitre}
\Gamma_0=\{X_{ij}(s)=X_{ji}^*(s),\quad 0\le s\le T\}
\eeq
In refs.~\cite{k1,ps} it was shown, using the Cauchy theorem,
that whenever $\calf$ is holomorphic in the matrix elements 
$X_{ij}$ one can modify the integration surface $\Gamma_0$
to the homotopically equivalent
\beq\label{ventiquattro}
\Gamma_1=\{\phi_{ij}(s)=\bar\phi_{ji}^*(s),
\; {\rm Im}\,\rho_i(s)=0,
\quad 0\le s\le T\}
\eeq
without affecting the value of the integral.
This means that in~\rf{ventuno} $\hat{\bar\phi}$ may be regarded as
the hermitian conjugate of $\hat{\phi}$.

We would like to remark that expressions similar to~\rf{quindici}
were obtained in the framework of conformal field theory~\cite{mor}.
However, the explicit form of the variables $\bar\phi$ and of the Jacobian
$\calj$, which are essential for any physical application of our method, were 
not computed there.

\section{The Lyapunov Spectrum}

We shall now define the Lyapunov exponents $\lambda_j$,
$j=1,\ldots,N$, according to the relation~\cite{vulp}
\beq\label{venticinque}
\lambda_1+\cdots+\lambda_k={1\over T}
\log\hbox{Vol}(\hat{P}_T\bfv_1,\ldots,\hat{P}_T\bfv_k)
\eeq
where $\bfv_1,\ldots,\bfv_k$ is an orthonormal set of vectors generating
a unitary $k$-volume.
For the sake of definiteness we shall choose $\bfv_j=\bfe_j$,
where $(\bfe_j)_i=\delta_{ij}$.
One has
\beq\label{ventisei}
\hbox{Vol}(\hat{P}_T\bfe_1,\ldots,\hat{P}_T\bfe_k)
=\sqrt{\sum_{a=1}^{l_k}\Delta_a^2(\hat{M}_k)}
\eeq
where $\Delta_a(\hat{M}_k)$, $a=1,\ldots,l_k$, 
$l_k\equiv\left(\matrix{n\cr k}\right)$
are the $k\times k$ minors of the $n\times k$ matrix
$\hat{M}_k=[\hat{P}_T\bfe_1,\ldots,\hat{P}_T\bfe_k]$.
Let $\bfphi_j,\bfp_j$, $j=1,\ldots,N$ be the vectors defined by
$(\bfphi_j)_i=\delta_{ij}+\phi_{ij}$,
$(\bfp_j)_i=(\hat{P}_T)_{ij}$.
Then
\bea
\nonumber
\bfp_j	&=&	
\sum_{i,k}
\left(\delta_{ik}+\phi_{ik}\right)e^{\tau_k}
\left(\delta_{kj}+\theta_{kj}\right)\bfe_i
\\
\label{ventisette}
&=&	\sum_{k=1}^N e^{\tau_k}\left(\delta_{kj}+\theta_{kj}\right)\bfphi_k 
\\
\nonumber
&=&	e^\tau_j\bfphi_j+\sum_{k<j}\theta_{kj}e^{\tau_k}\bfphi_k
\eea
From~\rf{ventisette} and the multilinearity of determinants it follows
\bea
\nonumber
\Delta_a(M_k)	&=&	\Delta_a[\bfp_1,\ldots,\bfp_n]	\\
\label{ventotto}
&=&	\Delta_a[e^{\tau_1}\bfphi_1,\ldots,e^{\tau_k}\bfphi_k]
\\
\nonumber
&=&	e^{\tau_1+\cdots+\tau_l}
	\left(
		\delta_{a,1}+\sum_{r_{ij}\ge 1}c_{ij}^{a,k}\phi_{ij}^{r_{ij}}
	\right)
\eea
where $\Delta_1$ is the minor obtained from the first $k$ rows of $M_k$,
$r_{ij}$ are strictly positive integers, and $c_{ij}^{a,k}$ are integer
coefficients with $c^{1,k}_{ij}\equiv 0$.
One has then
\bea\label{ventinove}
\lambda_1+\cdots+\lambda_k	&=&
{1\over T}[\tau_1(T)+\cdots+\tau_k(T))]
\\
&+& {1\over 2T} \log
\left[
	1+\sum_{a=2}^{l_k}
	\left(
		\sum_{r_{ij}\ge 1}c_{ij}^{a,k}\phi_{ij}^{r_{ij}}(T)		
	\right)^2
\right]
\nonumber
\eea
and
\beq\label{trenta}
\lambda_k={1\over T}\int_0^T\rho_k\,dt+
{1\over 2T}[\log(1+f_k(\hat{\phi}))-\log(1+f_{k-1}(\hat{\phi}))]
\eeq
where $1+f_k(\hat{\phi})$ is the argument of the logarithm in~\rf{ventinove}.

Let us now compute the probability distribution function for $\lambda_k$.
The form of~\rf{ventidue} implies that the $\phi$-dependent terms 
in~\rf{trenta} give no contribution, since they do not contain the
conjugate variables $\bar\phi_{ij}$.

We are therefore left with $N-1$ Gaussian integrations over 
$\rho_1,\ldots,\rho_{k-1},\rho_{k+1},\ldots,\rho_N$
which give the following exact result for the statistics of $\rho_k$:
\beq\label{trentuno}
\de\rho_k\,\expo{-{N\over 4D(N-1)}\int_0^T(\rho_k(s)-\bar\lambda_k)^2\,ds}
\eeq
where
\beq\label{trentadue}
\bar\lambda_k=D(N-2k+1),	\qquad	k=1,\ldots,N
\eeq
The probability distribution function $p(\lambda_k;T)$ 
of the $k$-th Lyapunov exponent $\lambda_k$ is then:
\beq\label{trentatre}
p(\lambda_k;T)={1\over 2}
\sqrt{NT\over\pi D (N-1)}
\expo{-{NT\over 4D(N-1)}(\lambda_k-\bar\lambda_k)^2}
\eeq
The Lyapunov exponents $\lambda_k$ are statistically
dependent due to the constraint~\rf{somma} and
their joint distribution function has a generalized 
Gaussian form~\footnote{
The Gaussian distribution of the Lyapunov exponent in
the $N=2$ case 
was obtained 
in the context of the passive scalar problem
in ref.~\cite{ps1}.
}.
We have thus 
obtained a complete knowledge of the statistics of the Lyapunov
spectrum of the matrix $\hat{P}_t$.
This has an essential application to the problem of
the computation of the multipoint correlators of a
passive scalar advected by a random velocity field
(see the Appendix).

\section{Conclusions}

In this work we gave a detailed exposition in the general
$N\times N$ case of a functional integral method for the
averaging of time-ordered exponentials of random matrices
which has found several applications to the study of the
statistical properties of disordered physical systems
\cite{k1,alpha,beta,chgk,ps1,ps,ps2} and
we have shown how the statistics
of the Lyapunov spectrum of a linear evolutionary process
can be computed exactly.
As a matter of fact,
our method provides a general setting for computing  the 
statistics of linear evolutionary systems subjected to
a white noise force field.

We would like to conclude with some remarks. 
The definition of the Lyapunov exponents as the
logarithmic rate of growth of a $k$-dimensional
parallelepiped (see eq.~\rf{venticinque} and ref.~\cite{vulp})
is the most natural from a physical point of view,
e.g. in the passive scalar problem.
Generally speaking these exponents do not coincide
with the logarithms of the eigenvalues of the evolution
matrix $\hat{P}_t$. The statistics of the
eigenvalues of a similar evolution matrix was
studied in refs.~\cite{dor,mpk,br,cas}.
Our method, however, allows one to obtain a more
detailed statistical information about the evolution
of initial vectors and to compute non-trivial
correlation functions of their components.
For an application to the passive scalar problem
see ref.~\cite{ps2}.

Lastly, we would like to remark that a more refined
application of the functional integral method we described
allowed
to solve effectively 
the more difficult case of a ``coloured'' noise
(see ref.~\cite{ps}).

\vspace{.6cm}

\begin{center}
\bf Acknowledgments
\end{center}

\noindent
We gratefully acknowledge inspiring discussions with P.~Casati, A.~Cattaneo,
M.~Chertkov, B.~Chirikov, G.~Jona-Lasinio, 
F.~Magri, M.~Martellini,
O.~Ragnisco, P.~Santini and A.~Vul\-pia\-ni. 
One of us (A.G.) would like to thank the Physics Department
of Rome University {\it La Sapienza}, where part of this
work was done, for warm hospitality.

\vspace{1cm}

\begin{center}
{\Large\bf Appendix}
\end{center}

\noindent
The method we exposed in this work has a direct application to
the computation of the statistics of a scalar passively advected
by a random velocity field.
In order to illustrate this point we will briefly recall here
the terms of the problem. For more detail see
refs.~\cite{fl,ps1,ps,chgk,ps2}.

The evolution of a scalar field $\theta(\bfr,t)$ passively advected by
a velocity field $\bfv(\bfr,t)$ and generated by a source 
$\phi(\bfr,t)$ is given by
\beq\label{eqa}
\dot\theta+\bfv\cdot\nabla\theta=\phi
\eeq
If we impose on~\rf{eqa} the asymptotic condition
$\theta(\bfr,-\infty)=0$
we get the solution
\beq\label{eqb}
\theta(\bfr,t)=\int^{+\infty}_0\phi(\bfR(\bfr,t-s),t-s)\,ds
\eeq
saying that
$\theta(\bfr,t)$ is completely determined in terms of
the trajectories $\bfR(\bfr_0,t)$ of the fluid particles:
\beq\label{eqc}
\dot\bfR=\bfv(\bfR,t),\qquad\bfR(\bfr_0,0)=\bfr_0
\eeq
Let us now take $\phi$ and $\bfv$ to be random, $\delta$-correlated
in time fields . 
The source $\phi$ will be assumed to be spatially correlated on a scale $L$. 
%
%
The velocity field $\bfv$ might be multi-scale, with smallest scale larger
or of the order of $L$.
The statistics of $\phi$ and $\bfv$ will be assumed to be spatially
isotropic.

Generally speaking, one is interested in computing equal-time correlators
of the form 
$\la\theta(\bfr_1,0)\theta(\bfr_2,0)\ra$
for 
$|\bfr_2-\bfr_1|\ll L$.
From the isotropicity of the statistics of $\phi$ and $\bfv$
it follows that such quantities are rotation-invariant.
Moreover,~\rf{eqb} 
implies that the statistics of $\theta$ is completely determined
in terms of the statistics of the trajectories~\rf{eqc}.

In order to subtract the effect or sweeping, let us chose a reference frame 
locally comoving with one of the fluid particles (see refs.~\cite{fl,ps1,ps}).
We can then locally linearize~\rf{eqc}, obtaining
\beq\label{eqd}
\dot\bfR\simeq\hat{\sigma}(t)\bfR
\eeq
where $\sigma_{ij}\equiv\partial v_j/\partial r_i$,
the matrix of velocity derivatives, will be taken to be
a random gaussian process. 
In the general case we have
$\hat{\sigma}=\hat{R}+\hat{S}$, where $\hat{R}$ 
is the antisymmetric part of $\hat{\sigma}$, inducing a rotation of 
the passive scalar blob, and $\hat{S}$ is the symmetric part, representing
the stretching of the unit blob.
We will consider the case of an incompressible fluid, so 
$\tr\hat{S}=0$.

More specifically, let us consider the following statistics of $\hat{\sigma}$:
\bea\label{eqe}
&&\de{\cal M}[\hat{\sigma}]=\de\hat{\sigma}
\,\expo{-{1\over 2 D_s}\int_0^T {\cal L}\,dt}\, ,
\\
\nonumber
&&
{\cal L}={1\over 2}
\left(\tr\hat{S}^2-{D_s\over D_r}\tr\hat{R}^2\right)
={1\over 2}
\tr\left(\hat{S}+i\sqrt{D_s\over D_r}\hat{R}\right)^2
\eea
Since we are interested in
rotation-invariant quantities
the final result shall be independent on $D_r$.
This arbitrariness allows one to substitute
$\sqrt{D_s/D_r}\rightarrow -i$,
$\hat{R}\rightarrow i\hat{R}$,
$\hat{\sigma}\rightarrow\hat{X}=\hat{S}+i\hat{R}$
(see ref.~\cite{chgk}),
and thus to consider a generic traceless hermitian 
matrix $\hat{X}$ with averaging weight
$\expo{-{1\over 2D_s}\int {1\over 2}\tr\hat{X}^2}$
in the place of the generic traceless real matrix $\hat{\sigma}$
with the averaging weight~\rf{eqe}: 
this allows one to refer to the results of sect.~2.

From rotational invariance it follows that the statistics
of $\theta$ is completely determined in terms of the statistics of the
Lyapunov spectrum of the matrix $\hat{P}_t$ defined by
\beq\label{eqf}
\dot{\hat{P}}_t=\hat{X}\hat{P}_t,\qquad\hat{P}_0=\one
\eeq
This reduces us to the problem studied in the preceding 
sections.
The Gaussian statistics of the Lyapunov exponents 
agrees with an old result~\cite{kr} about the Gaussian statistics
of a line element in a $\delta$-correlated in time velocity field.

\end{document}